\documentclass[]{iopart} 
\usepackage{geometry} 
\usepackage{graphicx} 
\usepackage{amssymb} 
\usepackage{enumerate}
\usepackage{subfigure}

\begin{document} \title{Searching for binary coalescences with inspiral
templates: Detection and parameter estimation}

\author{Benjamin Farr${}^{1,2}$, Stephen Fairhurst${}^{1}$, B.S.
Sathyaprakash${}^{1}$}
\address{$^{1}$ School of Physics and Astronomy, Cardiff University, The
Parade, Cardiff, UK}
\address{$^{2}$ Department of Physics, Rochester Institute of Technology, 84
Lomb Memorial Drive, Rochester, NY 14623\\
Center for Computational Relativity and Gravitation and Department of
Physics, Rochester Institute of Technology, Rochester, NY 14623}

\begin{abstract}
There has been remarkable progress in numerical relativity recently.
This has led to the generation of gravitational waveform signals
covering what has been traditionally termed the three phases of the
coalescence of a compact binary - the inspiral, merger and ringdown.  In
this paper, we examine the usefulness of inspiral only templates for
both detection and parameter estimation of the full coalescence
waveforms generated by numerical relativity simulations. To this end, we
deploy as search templates waveforms based on the effective one-body
waveforms terminated at the light-ring as well as standard
post-Newtonian waveforms. We find that both of these are good for
detection of signals.  Parameter estimation is good at low masses, but
degrades as the mass of the binary system increases.
\end{abstract}
\maketitle

\section{Introduction}

Several ground-based interferometric detectors are now in operation to
detect gravitational waves. These include the Laser Interferometric
Gravitational-wave Observatory (LIGO) at two sites in Livingston and
Hanford, USA, and the Virgo detector in Cascina, Italy.  They have
recently completed a first science run at or close to design sensitivity
and are sensitive to gravitational waves from coalescing binaries at
distances of tens to hundreds of megaparsecs depending on the total
mass and the mass ratio of the system. The broadband sensitivity
(40-400 Hz) of these detectors makes it possible to search for binaries
with  a rather large range of component masses from one to hundreds of
solar masses.  This range of masses includes both the neutron star
binaries (which are known to exist) as well as neutron star-black hole
and double black hole binaries (of which we have no observational
evidence).

In this article we test the efficiency of inspiral waveforms for the
detection and parameter estimation of the full coalescence signal.  We
restrict our attention to two waveform families.  The first of the
families is the Fourier domain model, called TaylorF2 or SPA
\cite{sathya-91}, which is an analytical approximation to the Fourier
transform of the standard post-Newtonian (PN) \cite{Blanchet:2002av,3PN,BFIJ02,BDEI04}
waveform (i.e., TaylorT3) computed using the stationary phase
approximation.  The highest PN order available in the LIGO Scientific
Collaboration (LSC) code base \cite{lal} for this family is $(v/c)^7$
(i.e. 3.5PN order). A number of searches by the LSC for compact binaries
of low masses (i.e. $M<25\, M_\odot$) have used this model to second
post Newtonian order as optimal templates
\cite{Abbott:2003pj,Abbott:2005pe,Abbott:2007xi,Abbott:2008}. 

The second family we consider is the effective one-body (EOB)
\cite{Buonanno:1998gg} model at four PN\footnote{Post-Newtonian expansion
is currently known only to 3.5 PN order 
\cite{Blanchet:2002av,3PN,BFIJ02,BDEI04}; the unknown 4 PN coefficient is
determined by matching the EOB to numerical relativity waveforms 
\cite{Buonanno:2000ef,Buonanno:2006ui,Buonanno:1998gg,Damour:2007yf,Damour:2007vq,Damour:2008te}
(see below).} (i.e., $(v/c)^8$) order but
terminated at the light ring.  As discussed below, the EOB model
provides the full waveform, i.e., the inspiral, plunge, merger and
ringdown waveform. In particular, the merger-ringdown waveform is
obtained by stitching a superposition of quasi-normal modes to the
inspiral-plunge waveform at the EOB light-ring
\cite{Buonanno:2000ef,Buonanno:2006ui,Buonanno:1998gg,Damour:2007yf,Damour:2007vq,Damour:2008te}.
In this study, however, we use the EOB model without the ringdown
modes.%
\footnote{Throughout this article, we will use EOB to refer to the EOB
model truncated at light ring, and EOBNR to refer to the EOB waveform
including the ringdown modes, calibrated to numerical relativity
results.}
Thus, it captures some part of the coalescence signal, and is therefore
better suited to search for higher mass signals.  Here we will focus on
the efficiency of our template bank to capture coalescence signals with
the TaylorF2 and EOB models.  In Section \ref{sec:bbh}, we discuss in
greater detail the dynamics of binary black hole mergers and the PN and
EOB models.

We test the efficiency and parameter estimation accuracy of the searches
in two different ways.  First, in Section \ref{sec:bank_eff} we perform
a Monte--Carlo study of the efficiency of the TaylorF2 and EOB models to
detect the full waveforms.  Since the full waveform is not known over
the entire parameter space, we make use of the EOB waveform calibrated
to the numerical relativity results
\cite{Damour:2007yf,Damour:2007vq,Damour:2008te,Buonanno:2007pf,Boyle:2008ge}.
Then, in Section \ref{sec:ninja}, we perform a similar comparison making
use of waveforms generated numerically.  This study was performed on the
NINJA \cite{Aylott:2009ya} data set which comprised simulated data for the LIGO
and Virgo detectors with numerically obtained binary coalescence signals
added. The original numerical results for the NINJA numerical waveform
contributions are described in
\cite{Hannam:2007ik,Hannam:2007wf,% BAM_HHB
Tichy:2008du,Marronetti:2007wz,Tichy:2007hk,% BAM_FAU
Pollney:2007ss,Rezzolla:2007xa,% CCATIE
% Hahndol
% LazEv
% Lean
Vaishnav:2007nm,Hinder:2007qu,% MayaKranc
Buonanno:2006ui,Pretorius:2007jn,% PU
Boyle:2007ft,Scheel:2008rj,% SpEC
Etienne:2007hr% UIUC
} (where these are published results), the codes are described in
\cite{Brugmann:2008zz,Husa:2007hp,% BAM
Koppitz:2007ev,Pollney:2007ss,% CCATIE
Imbiriba:2004tp,vanMeter:2006vi,% Hahndol
Zlochower:2005bj,Campanelli:2005dd,% LazEv
Sperhake:2006cy,% Lean
Hinder:2007qu,% MayaKranc
Pretorius:2004jg,Pretorius:2005gq,% PU
Scheel:2006gg,% SpEC
Etienne:2007hr% UIUC
}.

\section{Binary black hole dynamics}
\label{sec:bbh}

The evolution of a black hole binary is driven by back-reaction due to
the emission of gravitational waves which causes the system to inspiral
and merge. Confident detection of the emitted signal is greatly
facilitated by an accurate understanding of the dynamics of the binary
and the shape of the emitted waveform during inspiral and merger.  The
early evolution of a binary can be well-modelled by the PN approximation
during which the system slowly inspirals on an adiabatic sequence of
quasi-circular orbits located at the (stable) minimum of the changing
effective potential. In fact, for most of its lifetime a binary black
hole can be accurately described by the balance of the rate of change of
the binding energy with the energy carried away to infinity by the
radiation as given by the quadrupole formula. In fact, many of the
ideas (effective potential, last stable orbit, etc.) that are relevant 
when the component masses are greatly separated are still very useful in 
analytically modelling the system close to coalescence. However, they 
are perhaps not so useful or needed from the view point of numerical
evolution.

As the system evolves, the effective potential changes and reaches a
point when the potential transforms from one having a stable minimum and
an unstable maximum to one having just an unstable minimum. After this,
the system no longer possesses any bound orbits.  The transition point,
called the last stable orbit\footnote{The LSO discussed in this paragraph
refers to the test-particle limit of a binary when the mass ratio is
very small; PN corrections and resummed models modify the location of the 
LSO \cite{DIS1}. However, these changes are unimportant to the present discussion
where we are concerned with approximate numbers to determine the rough
boundary between where a certain phase of the evolution is dominant.} 
(LSO), occurs when the radius $r$ of the
orbit (in Schwarzschild coordinates) approaches $r \sim 6GM/c^2,$ where $M$
is the total mass of the system.  In terms of the dominant component of
the emitted radiation, this corresponds to a gravitational-wave
frequency of $f_{\rm insp} \simeq 440\, {\rm Hz} (M/10 M_\odot )^{-1}.$
Therefore, for masses less than about $10\,M_\odot$, only the inspiral
stage of the coalescence lies in the detector's sensitive band of 40-400
Hz. 

Once the system passes the LSO, the two black holes plunge towards each
other and merge in about one orbital time scale of the LSO to form one
single distorted black hole. This is the so-called merger phase which is
amenable to analytic description by a clever re-summation of the PN
approximation but more recently numerical relativity simulations have
provided a better understanding of the merger phase and continue to
provide new insights. The frequency of the waves during this phase
changes rapidly from $f_{\rm merge} \simeq 440\, {\rm Hz}(M/10\,
M_\odot)^{-1}$ to $1200\, {\rm Hz}\, (M/10\, M_\odot)^{-1}.$ During the
late stages of the coalescence, the highly distorted black hole, that
results from the merger of the two parent black holes, settles down to
an axi-symmetric quiescent state by emitting its deformation in the form
of gravitational waves. The radiation from this phase is well described
by black hole perturbation theory and consists of a set of quasi-normal
modes (often referred to as ringdown signal) whose fundamental frequency
is $f_{\rm ring} \sim 1800\, (M/10\,M_\odot)^{-1}$ when two equal mass
non-spinning black holes merge to form a single black hole whose spin
magnitude is estimated to be $J/M^2 \simeq 0.7.$ The first two overtones
of this mode have frequencies of $\sim 1650\, \rm Hz$ and $\sim 1700\,
\rm Hz,$ for the same system.

\subsection{Search templates}

The foregoing discussion hints that binaries whose total mass is less
than about $10\,M_\odot$ can be detected by using templates that are
described by the PN approximation. In fact, experience suggests that we
could make do with the PN waveform as templates even when the total mass
is as large as about $25\,M_\odot$ and they have been used in the search
for low-mass systems (i.e. systems with their total mass less than about
$25 M_\odot$) in the data from LIGO and Virgo
\cite{Collaboration:2009tt}. However, for higher mass black hole
binaries (i.e. binaries with their total mass greater than about $25
M_\odot$) the merger of the binary occurs in the detector's sensitive
band.  At merger, the dynamics is no longer adiabatic and is, therefore,
not well-modelled by PN expansion.  It has been a long standing aim of
numerical relativity to generate the full waveforms for gravitational
wave detection from higher mass black holes.

There has been significant progress recently in numerical relativity
with several groups having successfully simulated the merger of two
black holes (see, for example,
\cite{Pretorius:2005gq,Campanelli:2005dd,Baker:2005vv}, and the NINJA
related numerical relativity results cited earlier).  The longest of
these simulations last for tens of orbits \cite{Boyle:2007ft}, and they
cover different mass ratios and are beginning to explore the space of
component masses with spins.  

Nearly a decade ago, analytical work by Buonanno and Damour
\cite{Buonanno:1998gg} extended the PN dynamics beyond the last stable
orbit to calculate the merger dynamics. This analytical method, called
EOB computes the dynamics up to the light ring of the effective potential
and the waveform can be computed for separations larger than about
$r\simeq 2.2\, M.$ In this work we have used the EOB waveform terminated
at the light ring as search templates.  The EOB formalism, however,
provides the full waveform, including the merger-ringdown portion which
is attached at the EOB light ring.  Moreover, the availability of
numerical relativity simulations has helped in fixing certain unknown
higher order (4PN) terms in the EOB model by fitting the analytical
waveform to numerical relativity.  The current implementation of the
full EOB model in LAL uses the 4PN order parameter obtained in
\cite{Buonanno:2007pf} by calibrating the EOB model to the Goddard-NASA
numerical simulations, and three ringdown modes (i.e., the fundamental
mode and two overtones).  We note that, more recently, the EOB model
has been further improved by calibrating it to more and more accurate
numerical simulations
\cite{Damour:2007yf,Damour:2007vq,Damour:2008te,Boyle:2008ge} which
will form the basis of future searches. In the current study, EOBNR is 
used to calibrate the efficiency of the inspiral models.

\begin{figure}[ht]
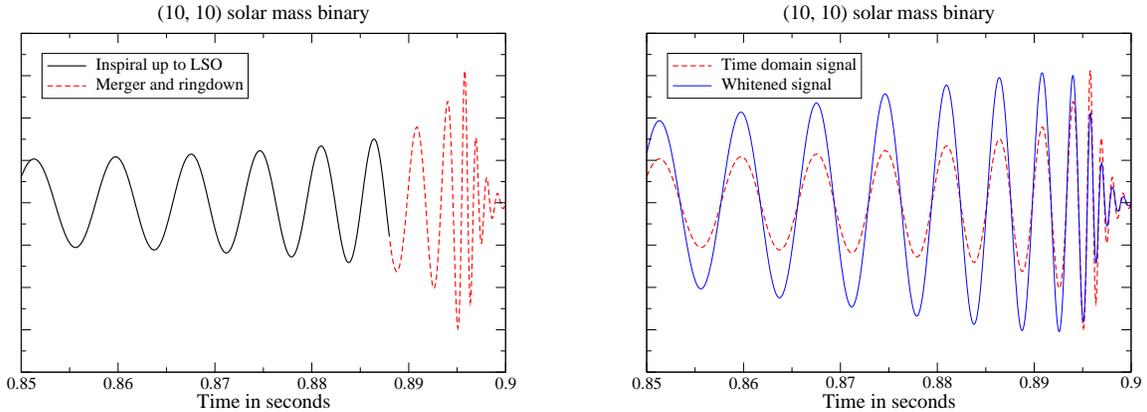

\centering
\vskip 1cm
\includegraphics[width=.45\textwidth]{Figures/fig1a}
\hfill
\includegraphics[width=.45\textwidth]{Figures/fig1b}
\caption{An example of the EOBNR waveform for a binary
consisting of two equal mass, non-spinning black holes each $10\, M_\odot.$
The EOB dynamics allows the computation of the inspiral (left panel,
black solid line) and plunge (left panel, roughly the first 
two cycles of the red dashed line) phases but the merger and ringdown waveform
(left panel, latter part of the red dashed line) is stitched to the end of the 
plunge phase by matching the amplitude of the waveform and its first 
two derivatives by using a superposition of the fundamental quasi-normal 
mode and its first two overtones. The right panel compares the 
time-domain signal (red dashed line) with the whitened signal 
(blue solid line). }
\label{fig:EOBNR}
\end{figure}

Ajith et al \cite{Ajith:2007kx,Ajith:2007qp} have used a
phenomenological approach to match the inspiral phase from PN
approximation to the merger and ringdown from numerical relativity.
Here, an hybrid waveform obtained by stiching together the PN inspiral
waveform to the numerical-relativity inspiral, merger and ringdown
waveform is first built. Then, the phenomenological waveform in the
frequency domain is constructed by requiring closeness to the hybrid
waveform.  In the long run, it is likely that these full waveforms will
be used as templates to search for inspiral signals in
gravitational wave detectors.  

%% At present, however, phenomenological 
%% waveforms describing the full signal in the case where the
%% component objects have large spins are still under development.

In order to test the efficiency of EOB and TaylorF2 families, we will use 
the EOBNR waveform as our ``true'' waveform and see how well these 
partial waveforms perform in both detection and parameter estimation.

\subsection{An example waveform}

Fig.\ \ref{fig:EOBNR} shows the waveform expected from a pair of
non-spinning $10\,M_\odot$ black holes during the last 50 ms before
merger. The left panel shows the time-domain waveform $h(t)$ and the
right panel compares the time-domain waveform with the signal
`perceived' by the initial LIGO detector whose noise has been whitened.
In other words, the right panel plots whitened template $q(t)$ given 
by $$q(t) = \int_{-\infty}^\infty \frac{H(f)}{\sqrt{S_h(|f|)}} \, 
\exp(-2\pi ift)\, {\rm d}f,$$ 
where $H(f)=\int_{-\infty}^\infty h(t)\, \exp(2\pi ift)\, {\rm d}t$ is the
Fourier transform of the time-domain signal and $S_h(f)$ is the
one-sided noise power spectral density of initial LIGO.  In the left
panel, the inspiral part of the signal given by EOB dynamics is shown in
black solid line followed by the plunge, merger and ringdown phases in red
dashed line.  The right panel compares the time-domain waveform in the
left panel (red dashed line) with the whitened signal (blue solid line).

Note that although the time-domain signal is dominated by the plunge,
merger and the ringdown phases, the detector noise spectral density (i.e, $S_h(f)$) 
suppresses them, making the inspiral phase more dominant. For systems 
with greater masses, more of the merger phase appears in band. 
For systems with total mass larger than about $80\,M_\odot$ 
the merger and ringdown signals begin to dominate over the inspiral 
phase. For such systems it is important to deploy EOBNR
templates. As a result, we cannot expect our template families 
to do well in capturing high mass binaries.

\section{Bank Efficiency}
\label{sec:bank_eff}

Matched filtering, the data analysis technique used in most searches for
binary black holes, is pretty sensitive to the phasing of the signal,
which in turn depends on (a subset of) the source parameters.  In the
case of non-spinning black holes on a quasi-circular orbit, the only
parameters that we must consider are the two masses.  The location of
the binary on the sky, the distance to the source, the polarization of
the wave, etc., are not important for a single detector as they simply
affect the amplitude of the waveform.  Although we won't know the time at
which the binary merges nor the phase of the signal at that epoch, these
parameters need not be explicitly searched for \cite{sathya-91}, and are
easily extracted in the process of maximising the cross-correlation of the
template with the data.

\subsection{Template bank}

Our goal in this Section is to study the efficiency of the two template
families in detecting binary black hole coalescences. To this end we
first set up a template bank --- a set of points in the parameter space
of the component masses.  A geometric algorithm described by Babak et al
\cite{Babak:2006ty} is used to generate the template bank and it is the
same algorithm irrespective of which family of waveforms is used to
filter the signals.  The bank is designed to cover the desired range
of component masses of the binary.  In addition to the range of the
component masses, our template bank algorithm requires us to specify a
parameter called the {\em minimal match,} $MM.$ The minimal match is the
smallest overlap guaranteed between a signal with random source
parameters and the template nearest (in the geometrical sense) to it in
the parameter space. 

Assumptions made in the construction of the
template placement algorithm imply that this will be strictly true only
when (a) the templates and signals belong to the same family, and (b)
the ending frequency (i.e., the LSO or light ring depending on the
waveform in question) is greater than the upper end of the sensitivity
band. The latter condition further implies that we can hope to achieve
overlaps of $MM$ or greater only for waveforms whose total mass is
smaller than a certain value depending on the detector bandwidth; in the
case of initial LIGO this is $10\,M_\odot.$ We have chosen $MM=0.95.$ However,
since our templates and signals belong to different PN approximations we 
cannot expect to achieve this overlap even for total mass less than 
$10\,M_\odot.$ 

The template placement algorithm chooses a hexagonal
grid in the two-dimensional parameter space of the component masses and
it is an optimal algorithm in the sense that it gives the smallest
number of templates possible for a given minimal match
\cite{Cokelaer:2007kx}.

Having constructed a template bank we then generate an EOBNR signal with
the values of its masses, the epoch and the phase at coalescence,
all chosen randomly but in a given range. Since the EOBNR extends beyond
the LSO, it is possible to generate signals with the total mass in the
range $[10,\,300]\, M_\odot$ and minimum component mass of $5\,M_\odot.$
We performed simulations for both the TaylorF2 and EOB templates over a
restricted range of the parameter space and an additional EOB analysis
over a much larger range.  Specifically, in the case of TaylorF2 we set
up a template bank with the total mass in the range $[6,\,80]\,M_\odot$,
with minimum component mass of $3 M_{\odot}$, and searched for signals
with total mass in the range $[10,\,72]\,M_\odot$.  Although the
TaylorF2 model is not expected to be a good approximation above about
$\sim 25\,M_\odot$, we are using it to much higher mass.  For the EOB
family truncated at light ring, we performed two separate analyses.  The
first had templates covering the range $[6,\,120]\,M_\odot$, minimum
component mass $3 M_{\odot}$ and injected signals with total mass in the range
$[20,\,80]\,M_\odot$.  Since the EOB waveforms extend to a significantly
higher frequency than the standard post Newtonian ones, we also
performed an analysis with templates of total mass in the range
$[60,\,400]\,M_\odot$ and minimum component mass of $30 M_{\odot}$.  In
this case the template bank consisted of only 9 templates.  The
simulated signals had total mass between $70$ and $300 M_{\odot}$.

Next, for each point in the template bank we generate waveforms from our
template families (EOB and TaylorF2) and measure their overlap with the
random signal. The overlap ${\cal O}_k$ of the $k$th template $q_k(t;\,
m_1^k,\, m_2^k)$ and the signal $h(t)$ is defined by 
$${\cal O}_k(m_1^k,\, m_2^k) = \max_t
2 \int_{f_l}^{f_u} \left [ H(f) Q_k^*(f;\, m_1^k,\, m_2^k) +
H^*(f) Q_k(f;\, m_1^k,\, m_2^k) \right ] \, e^{-2\pi ift}\, 
\frac{{\rm d}f} {S_h(f)},$$
where $H(f)$ and $Q(f)$ are the Fourier transforms of $h(t)$ and $q(t),$
respectively, and $Q^*$ is the complex conjugate of $Q.$ This allows us
to compute the maximum overlap between our template waveforms and a
random signal\footnote{For the sake of saving space we have not
discussed the maximization over the phase of the signal. This can be
found, for instance, in Ref.\ \cite{sathya-91}.}.  This process is
repeated for 1,000 different realizations of the random mass parameters
and the maximum of the overlap over the entire template bank is recorded
in each case.  We will now discuss the results of these simulations.

\subsection{Efficiency for detection}
\label{sec:det eff}

\begin{figure}[ht]
\includegraphics[width=.49\textwidth]{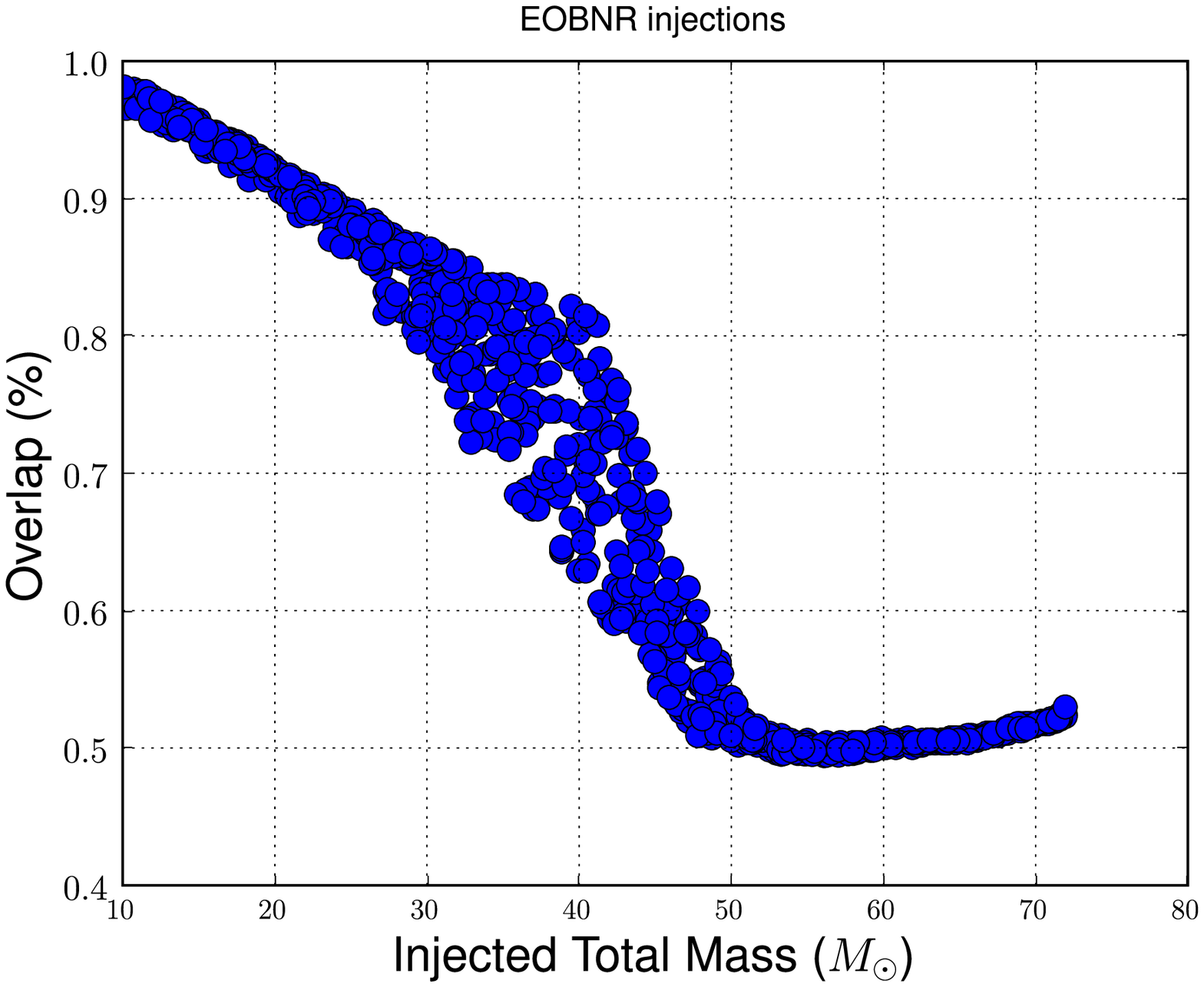}
\includegraphics[width=.49\textwidth]{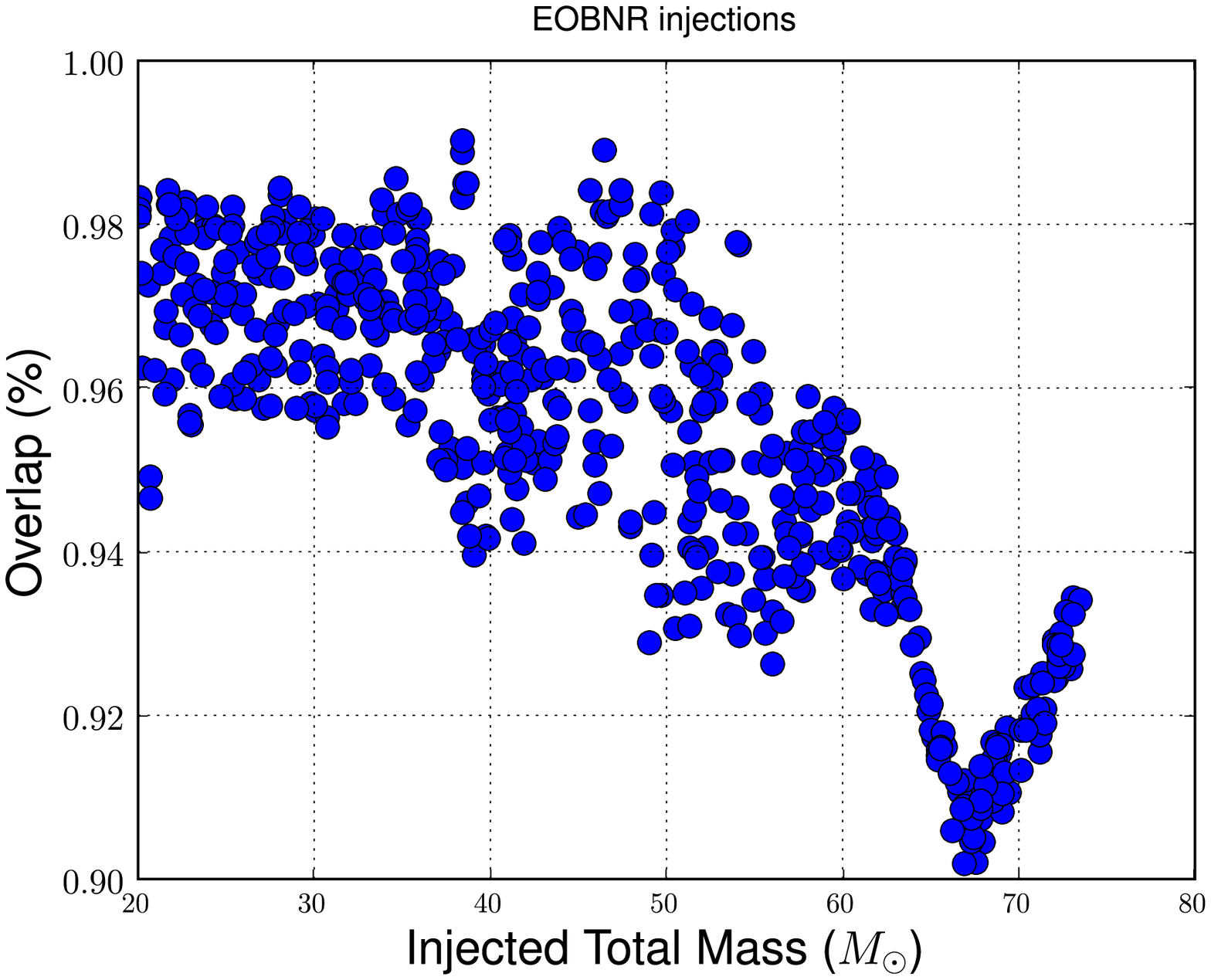}
\caption{These plots depict the efficiency of the TaylorF2 (left
panel) and EOB (right panel) template banks in detecting the 
coalescence waveforms assumed to be well-represented by EOBNR. Each
dot corresponds to the overlap of a random EOBNR signal maximized
over the template bank consisting of TaylorF2 or EOB waveforms.}
\label{fig:TBE}
\end{figure}

Fig.\ \ref{fig:TBE} plots the results of our simulation. The left panel
shows the (maximum) overlaps of the TaylorF2 template bank with random
signals, one dot for each trial. The right panel shows the same but for
the EOB template bank. The TaylorF2 model has overlaps larger than 90\%
for only signals whose total mass is less than about $22\,M_\odot.$ The
overlap falls off quickly for masses larger than this, reaching slightly
more than 0.5 when the total mass is about $50\,M_\odot.$ The fact that
the overlap remains unchanged beyond $50\,M_\odot$ is probably due to
the TaylorF2 template matching the later merger or ringdowm part of the
EOBNR waveform.  In contrast, the EOB model achieves an overlap greater
than $90\%$ over the entire mass range.  This is due to the fact that
the EOB waveform extends to the light ring.  

\begin{figure}[ht]
\centering
\includegraphics[width=.49\textwidth]{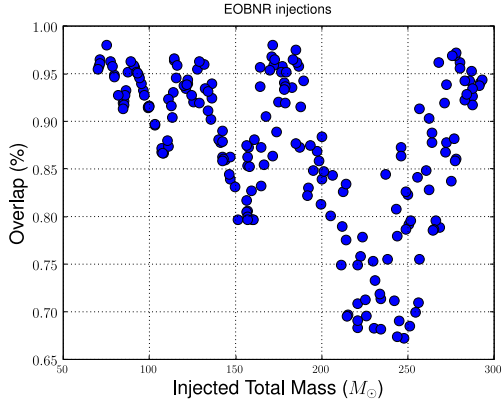}
\caption{This plots depict the efficiency of the EOB template bank
generated for the mass range $[60,\,400]\,M_\odot$ in detecting the 
coalescence waveforms assumed to be well-represented by EOBNR. Each
dot corresponds to the overlap of a random EOBNR signal maximized
over the template bank consisting EOB waveforms.}
\label{fig:BE_high_mass}
\end{figure}

Figure \ref{fig:BE_high_mass} shows the overlap obtained using the high
mass EOB template bank and an extended range of EOBNR simulated
waveforms.  We obtain overlaps of better than $85\%$ for systems whose
total mass is less than $150\,M_\odot,$ and the overlaps remain more
than $65\%$ even for systems with total mass less than $200\,M_\odot.$
The oscillatory behaviour seen in this case is due to an edge effect:
the density of templates gets smaller and smaller as the total mass
becomes larger. For this range of the total mass, there are only a 
handful of templates in the entire bank;  as a result, a single 
template might be available for a pretty large range of masses, 
causing the overlaps to swing up and down as the total mass is increased.
 
The overlaps are surprisingly high, especially considering that there
are only nine templates in the bank.  It is possible that the very high
mass EOB signals are very short in the detector band (perhaps a cycle or
two) and the abrupt cutoff of the template as a result of termination at
the light ring could bleed power into a frequency region where there is
no power in reality\footnote{The spurious power is in itself not a bad
thing but large noise glitches in the region where there is spurious
power could cause false alarms. This is especially the case when the
detector noise is contaminated by large amplitude non-stationary noise
glitches.}.  This spurious template power may lead to large overlaps
with the merger and ringdown parts of the EOBNR waveforms.  TaylorF2
does not suffer from this predicament. This is because TaylorF2 is
generated in the Fourier domain and the abrupt cutoff of the signal does
not cause any problem in the frequency domain and we are unconcerned
with spurious effects in the time-domian as they occur outside the
region of our interest.  

\subsection{Efficiency for parameter estimation}

The matched filtering statistic gives the likelihood for a signal to be
present in the data as opposed to the data being pure background noise.
The parameters of the template which maximize the likelihood are 
maximum likelihood estimates. Having determined the efficiency of our 
template banks in capturing the coalescence signals, we next consider
how good they are in measuring the signal parameters in the maximum
likelihood sense. If the template waveforms and the signal they are
intended to detect both belong to the same family then in the limit of
large signal-to-noise ratio the distribution of the maximum likelihood 
estimates will be centred on the true signal parameters. We are in a
situation wherein the template waveforms and the signals they are 
intended to recover belong to different approximations. Therefore, one
can expect a systematic bias in the estimation of parameters. 

\begin{figure}[ht]
\centering
\includegraphics[width=.49\textwidth]{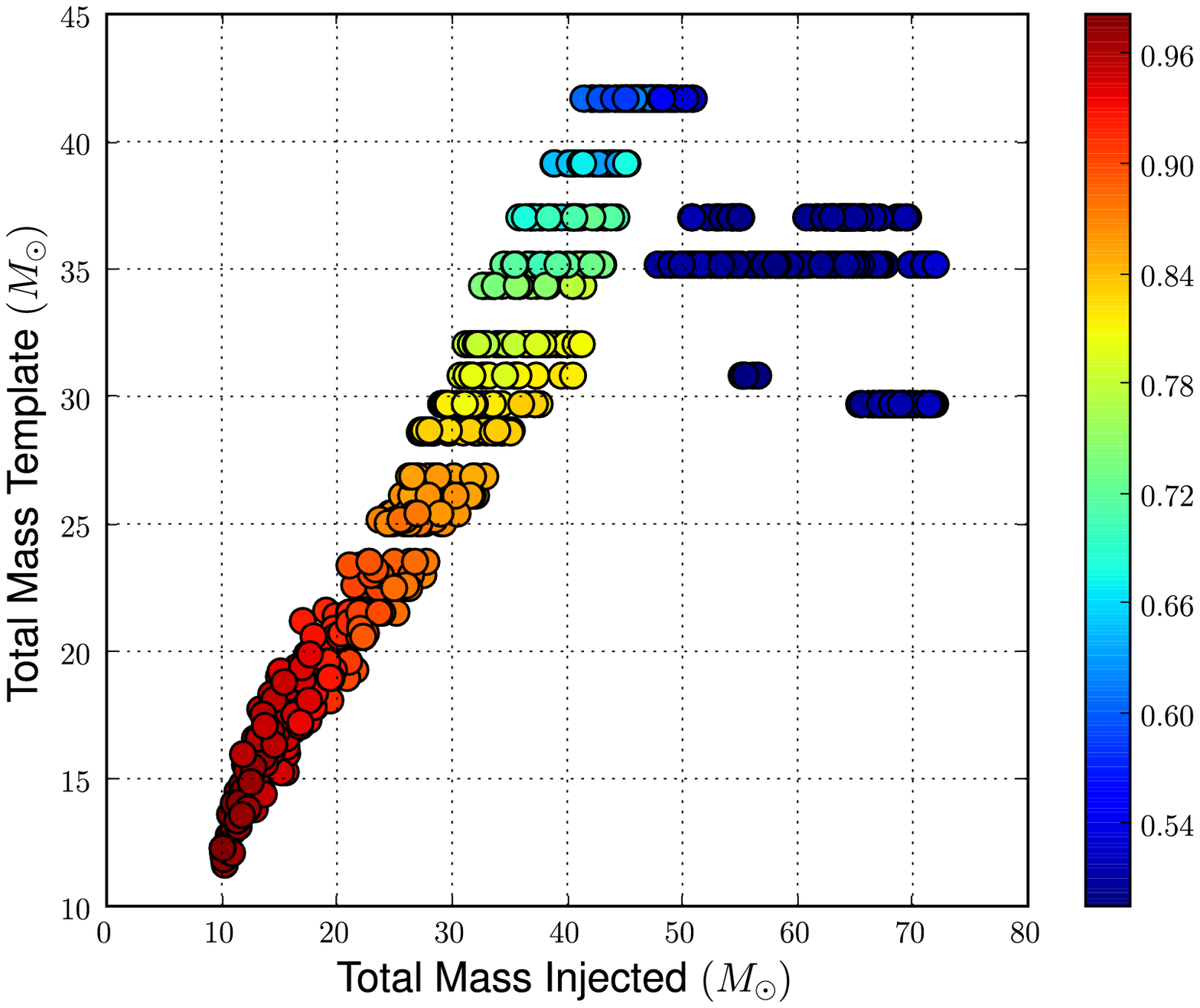}
\includegraphics[width=.49\textwidth]{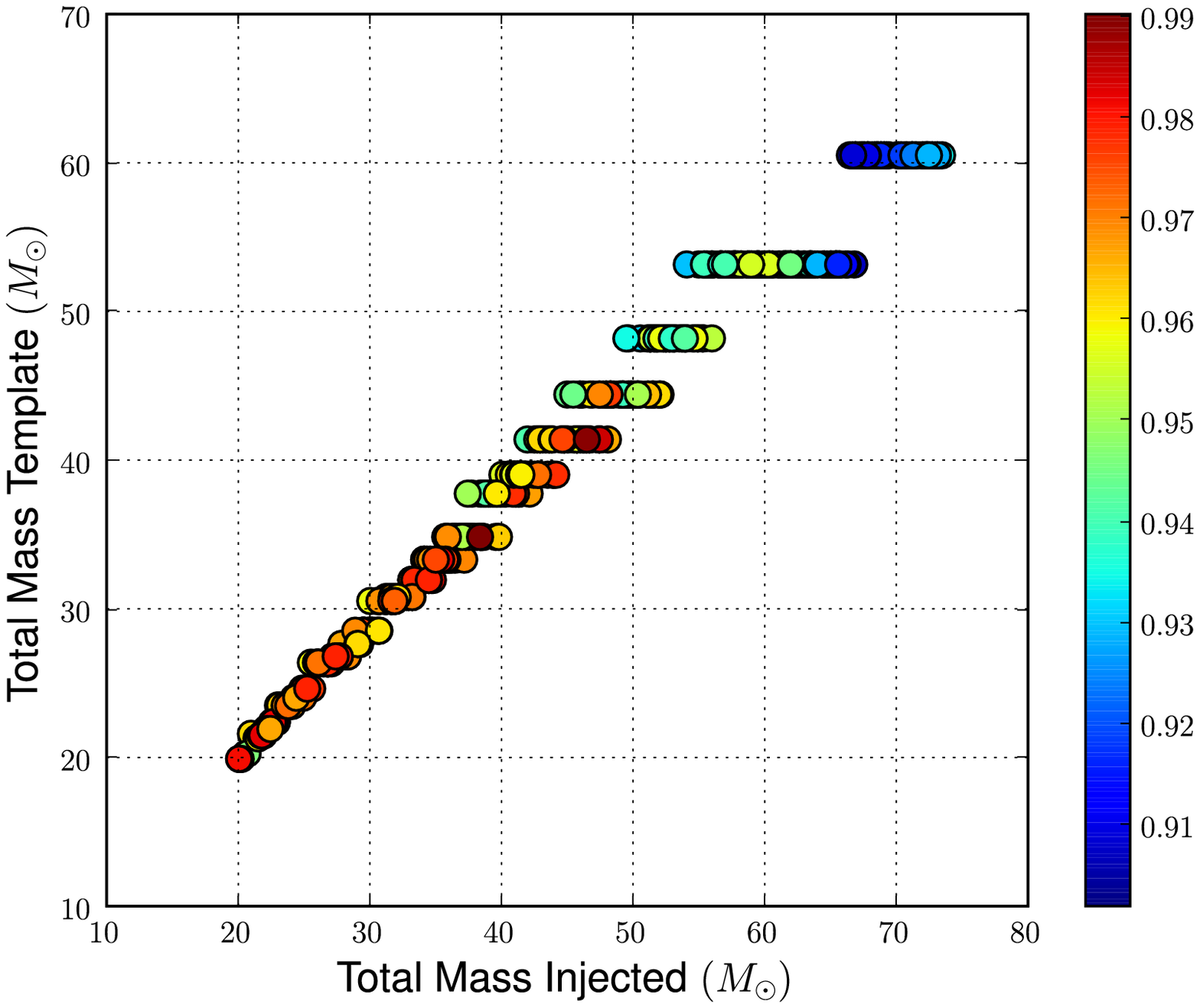}
\caption{Parameter estimation accuracies for the TaylorF2 (left panel)
and the EOB (right panel) models. The total mass of the injected EOBNR
signal is plotted vs the total mass of the template that obtained the
best overlap. The colour of the plotted point is determined by the
overlap between signal and template.  Clearly, there is a positive
correlation between the injected and measured masses in the case of
TaylorF2 model, with the spread in the measured values becoming larger
at higher masses.  Additionally, for higher masses the overlap
decreases.  Interestingly, all simulated signals with the total mass 
greater than $50 M_{\odot}$ are recovered with the same template of mass
$35 M_{\odot}$.  The EOB templates show a similar correlation between
simulated and recovered masses.  However, since the EOB templates extend
to higher frequency, the EOB templates are capable of accurately
recovering the higher mass simulations with accurate mass parameters and
an overlap greater than $90\%$.}
\label{fig:EBE}
\end{figure}

To gauge the reliability of the two families in estimating the parameters
of the true signal we make use of the results of the simulation from the
previous Section.  This simulation computed the overlap of the 
templates with the signals in the absence of any noise. Therefore, the  
parameters of templates that maximized the overlap when compared to the
true parameters of the EOBNR signal give a measure of
the systematic bias in parameter estimation due to the difference
in the waveform families representing the templates and the signal.
Fig.\ \ref{fig:EBE} shows the measured total mass of 
the template ($y$-axis) vs the true total mass of the EOBNR signal
($x$-axis) and the colour is determined by the corresponding overlap.

\begin{figure}[ht]
\centering
\includegraphics[width=.49\textwidth]{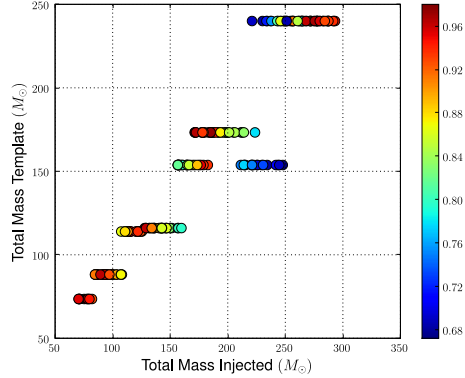}
\caption{Parameter estimation accuracies for high masses using the EOB
templates. The total mass of the injected EOBNR signal is plotted vs the
total mass of the template that obtained the best overlap. The colour of
the plotted point is determined by the overlap between signal and
template.  Even for high masses, there is a good correlation between the
simulated and recovered masses.  The discreteness of the EOB template
bank is clearly seen from the limited set of recovered masses, and
indeed this seems to account for the majority of poor overlaps.}    
\label{fig:EBE_high_mass}
\end{figure}

Both the TaylorF2 and EOB models provide good parameter recovery as well
as good overlap for simulated signals with a total mass less than $30
M_{\odot}$.  At masses higher than this, the TaylorF2 waveforms achieve
a significantly lower overlap, although parameter estimation remains
good up to a total mass of about $50 M_{\odot}$.  Above this mass, all
simulated signals are recovered by a template of mass $35 M_{\odot}$.
Interestingly, this template will terminate around $120 Hz$ which is
close to the most sensitive frequency of the LIGO detector.  Thus, it
seems likely that the inspiral template is picking up power from the
merger and ringdown parts of the EOBNR waveform.  For the EOB templates,
the overlap does drop off somewhat for higher masses, but the parameter
recovery remains reasonable throughout, with a slight tendency to
underestimate the total mass of the signal.  For higher masses, Figure
\ref{fig:EBE_high_mass} shows the parameter estimation accuracy for the
EOB waveforms.  Despite the fact that there are only a handful of
templates in the bank, the parameter recovery is reasonable.  

\section{NINJA Results}
\label{sec:ninja}

The Numerical INJection Analysis (NINJA) project was a mock data
challenge, where the data were generated at the design sensitivity of
the initial LIGO and Virgo detectors and numerical relativity waveforms
provided by a number of groups were added to the data.  A number of data
analysis methods were applied to the data, and the results of the NINJA
project are available elsewhere \cite{Aylott:2009ya}.  For the NINJA analysis,
we performed several runs through the data using the LSC's Compact
Binary Coalescence (CBC) analysis pipeline.  Here, we restrict our
attention to two runs through the data which are similar to the TaylorF2
and EOB analyses described in the previous Section.  This allows us to
investigate the issues of detection and parameter estimation using these
templates to search for full waveforms obtained from numerical
relativity.  The results are similar to those obtained in the previous
Section, namely that the EOB search has a greater efficiency than the
TaylorF2 search, but that both can detect high mass signals, although
the parameter estimation is poor.

The CBC pipeline was designed to analyze data from a network of
detectors to search specifically for gravitational wave signals from
binary neutron stars and black holes \cite{Abbott:2007xi}.  It proceeds
as follows: First, a bank of templates covering the desired mass range
is produced.  For the NINJA analysis we used a template bank covering
masses between $20$ and $90M_{\odot}$ with a minimal match of 0.97.  The
data from each of the detectors is separately match filtered against the
template waveforms \cite{Allen:2005fk}.  For this analysis, we
restricted attention to the mock data generated for the three LIGO
detectors (the 4 km and 2 km detectors denoted H1 and H2, respectively, at
Hanford, WA and the 4 km L1 detector at Livingston, LA).  A trigger is
produced whenever the signal-to-noise ratio exceeds the desired
threshold of $5.5$.  A coincident trigger is recorded whenever there are
triggers from two or more detectors with comparable masses and
coalescence times \cite{Robinson:2008un}.  Finally, these coincident
triggers are subjected to a set of signal based vetoes, in particular
the $\chi^{2}$ \cite{Allen:2004gu} and $r^2$ \cite{Rodriguez:2007}
tests, designed to separate signals from non-stationary transients in
the noise.  

\begin{figure}
    \includegraphics[width=0.5\linewidth]{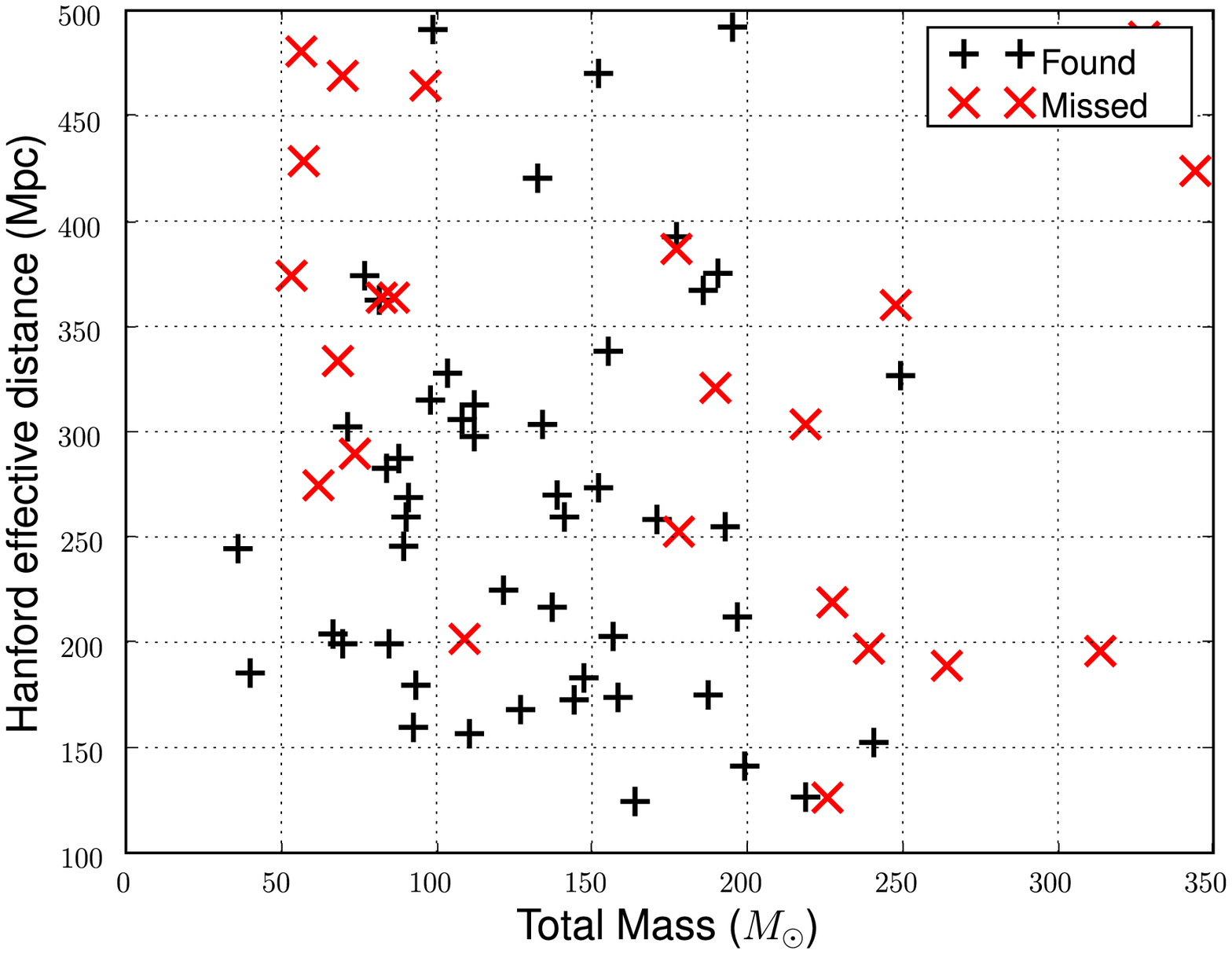}
    \includegraphics[width=0.5\linewidth]{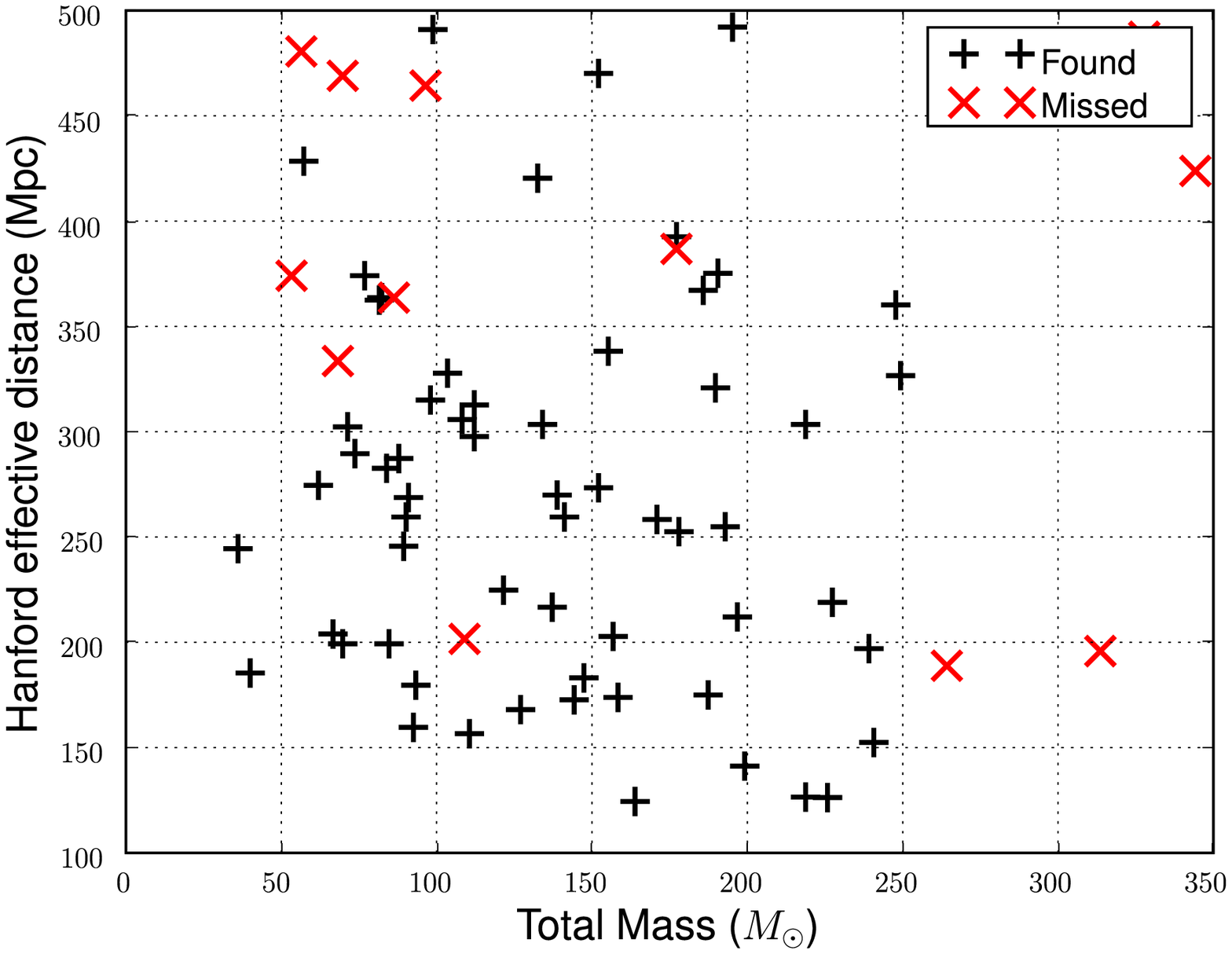}
\caption{Found and missed numerical injections for the TaylorF2 (left
panel) and EOB (right panel) searches of the NINJA data.  The found and
missed injections are plotted on the total mass, Hanford effective
distance plane.  The effective distance for a detector provides a
measure of the amplitude of the signal at that site, taking into account
the distance and orientation of the source.  For both searches, the
majority of the close simulations are recovered.  EOB templates are seen
to perform better, particularly at higher masses. In both searches,
the simulation with total mass $105 M_{\odot}$ and effective distance of
$200$Mpc is missed.  This has a peak amplitude at the
start of the waveform, rather than coalescence, and although a
coincident trigger is recorded the time between it and the waveform peak
is too large for our algorithm to associate them.}
\label{fig:found_missed}
\end{figure}

Our results are summarized in Table \ref{tab:cardiff_results} which
shows the number of injections recovered by the analysis pipeline at
each stage of the analysis for the two searches described here.  The EOB
search is capable of detecting a greater fraction of the simulated
signals than the TaylorF2 templates truncated at LSO.  This is further
highlighted in Figure \ref{fig:found_missed} where we show those
simulated signals which were recovered by the two different waveform
families.  The EOB model clearly performs better, particularly at higher
masses.  This is consistent with the findings of the previous Section. 

\begin{table}
\begin{center}
\begin{tabular}{| l || c | c |}
\hline
\bf{Template} & TaylorF2 & EOB \\ \hline
\bf{Freq. Cutoff} & LSO & Light Ring \\ \hline
\bf{PN Order} & 2 pN & 2 pN \\ \hline
\bf{Found Inj, Single Detector (H1, H2, L1)} & 72, 43, 66 & 91, 64, 82 \\ \hline
\bf{Found Inj, Coincidence} & 59 & 83 \\ \hline
\bf{Found Inj, Coincidence + Signal Vetoes} & 59 & 80\\ \hline
\end{tabular}
\end{center}
\caption{Results of inspiral search for NINJA waveforms.  There were 126
injections performed into the data.  The table above shows the number of
injections which were recovered using the two waveform families.  The
EOB search shows a significantly higher sensitivity than the TaylorF2
waveforms evolved to LSO.  Note that virtually all
simulations which pass the initial coincidence requirement also
survived the signal consistency checks.}
\label {tab:cardiff_results}
\end{table}

\begin{figure}
\begin{center}
\includegraphics[width=0.6\linewidth]{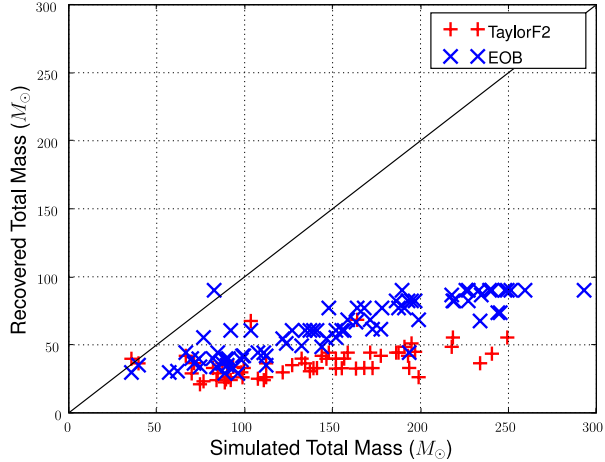}
\end{center}
\caption{Accuracy of recovering the total mass of simulated signals for
the TaylorF2 (red $+$) and EOB (blue $\times$) models.  For both of the
searches, the total mass is estimated poorly and systematically lower
than the simulated mass.  This is due to the fact that the search has
been performed with inspiral only waveforms for which the search extends
only up to $90 M_{\odot}$.}  
\label{fig:total_mass}
\end{figure}

Next, we turn to parameter estimation. Figure \ref{fig:total_mass} shows
the accuracy with which the total mass of the simulated signals is
recovered using the inspiral only waveforms.  For both the TaylorF2 and
EOB models, the parameter recovery is poor, particularly at higher
masses.  This is to be expected, since we are searching with partial
waveforms and, at the higher masses, it is the merger and ringdown of
the simulations which occupies the sensitive band of the detectors.
Furthermore, the template bank extends only to a total mass of $90
M_{\odot}$ making accurate parameter recovery of the high mass
simulations impossible.

\section{Discussion}

In the coming years, the first detection of gravitational waves from
coalescing binaries will surely be achieved.  Following the first
detection, attention will focus on extracting as much astrophysical
information as possible from the observed signal.  In the studies
described here, we have addressed the ability to perform both the
detection and parameter estimation problems using template waveforms
which cover only part of the binary coalescence.  We have made use of
two different waveforms --- the TaylorF2 PN waveforms taken to LSO and
the EOB waveforms to light ring.  In addition, we have used two
different methods to evaluate the detection and parameter estimation
capabilities of these signals --- a Monte--Carlo study using EOBNR
waveforms as the ``true'' signal, and an analysis of numerical
relativity waveforms in the NINJA data.  In all cases, the conclusion is
the same:  the inspiral only templates are useful for detection of the
signal, but do not provide good parameter estimation, particularly for
the higher mass signals.  This is to be expected as, for high mass
binaries, it is the merger and/or ringdown which occurs at the most
sensitive frequency of the detectors.  We observe that the EOB waveforms
perform somewhat better than the TaylorF2 waveforms.  This is expected
as the EOB waveforms extend to the light ring and therefore capture the
plunge part of the waveform which is not incorporated in the TaylorF2
PN model.

The results of this study show that post-Newtonian based inspiral only
waveforms will not be sufficient for satisfactory detection and
parameter estimation of higher mass black hole binaries.  Full waveforms
derived from a synthesis of post-Newtonian waveforms and numerical
relativity results, such as the EOBNR model
\cite{Damour:2007yf,Damour:2007vq,Damour:2008te,Buonanno:2007pf,Boyle:2008ge},
or phenomenological models, such as \cite{Ajith:2007kx,Ajith:2007qp},
will be necessary.

\section*{Acknowledgements}

We thank Duncan Brown, Alessandra Buonanno, Mark Hannam, Sascha Husa,
Badri Krishnan, Larne Pekowsky and Lucia Santamaria for helpful comments
on the analysis and paper.  We are greatful to Alessandra Buonanno, Evan
Ochsner and Craig Robinson for discussions on and coding up the EOBNR model.
We thank the Kavli Institute for Theoretical
Physics (KITP) Santa Barbara for hospitality during the workshop
``Interplay between Numerical Relativity and Data Analysis'', where the
NINJA project was initiated; the Kavli Institute is supported by NSF
grant PHY 05-51164.  BS acknowledges the support of the  UK Science and
Technology Facilities Council.  SF was funded by the Royal Society.  BF
was funded by the International REU program and grant  NSF-0838740.

\section*{References}
\bibliography{ninja}

\end{document}